# Inter-Operator Feedback in Data Stream Management Systems via Punctuation

Rafael J. Fernández-Moctezuma
Portland State University
Department of Computer Science
P.O. Box 751, Portland, OR 97207

rfernand@cs.pdx.edu

Kristin A. Tufte
Portland State University
Department of Computer Science
P.O. Box 751, Portland, OR 97207

tufte@cs.pdx.edu

Jin Li
Portland State University
Department of Computer Science
P.O. Box 751, Portland, OR 97207

jinli@cs.pdx.edu

## ABSTRACT
High-volume, high-speed data streams may overwhelm the capabilities of stream processing systems; techniques such as data prioritization, avoidance of unnecessary processing and on-demand result production may be necessary to reduce processing requirements. However, the dynamic nature of data streams, in terms of both rate and content, makes the application of such techniques challenging. Such techniques have been addressed in the context of static and centralized query optimization; however, they have not been fully addressed for data-stream management systems. In this work, we present a comprehensive framework designed to support prioritization, avoidance of unnecessary work, and on-demand result production over distributed, unreliable, bursty, disordered data sources, typical of many streams. We propose a form of inter-operator feedback, which flows against the stream direction, to communicate the information needed to enable execution of these techniques. This feedback leverages punctuations to describe the subsets of interest. We identify potential sources of feedback information, characterize new types of punctuation to support feedback, and describe the roles of producers, exploiters, and relayers of feedback that query operators may implement. We also present initial experimental observations using the NiagaraST data-stream system.

## Categories and Subject Descriptors
H.2.4 [**Systems**]: Query Processing

## 1. INTRODUCTION
Effective resource management is critical in data stream systems, but is complicated by the dynamic nature of data streams. We propose a new *feedback punctuation* mechanism that supports prioritized processing, avoidance of unnecessary work and on-demand result production in a data-driven fashion. Feedback punctuation enables more efficient utilization of data-stream-system resources.

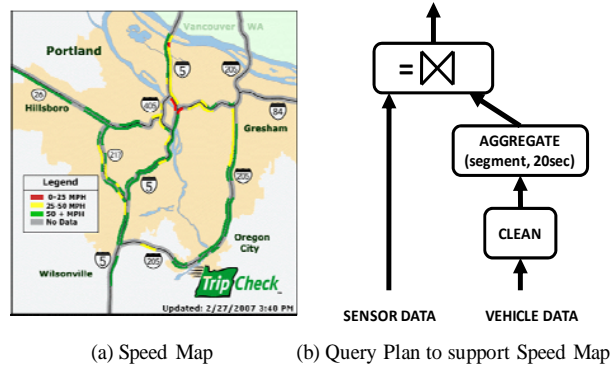

(a) Speed Map  (b) Query Plan to support Speed Map

**Figure 1. Speed Map for the Portland, OR Metro Area, and Query Plan to Support it. Image from Tripcheck.com**

Consider a traffic operations center (TOC) that produces a speed map of a metropolitan area, similar to the map shown in Figure 1 (a). Note that the freeway network is broken into short freeway segments and the speed map displays color-coded speeds for each segment. The TOC has two sources of data: a stream of data from a set of fixed sensors in the road network and a stream of readings from vehicles with on-board GPS systems. Fixed sensors provide traffic speed and volume data for their locations; vehicle data consists of speed and location for each vehicle. The TOC has decided that the fixed-sensor readings are adequate for estimating speed on uncongested freeway segments. However, for congested segments, vehicle readings will be used in addition to fixed-sensor readings to estimate traffic speed. Figure 1 (b) shows a possible plan for this query. The sensor data stream contains one report for each segment every 20 seconds. As there are many vehicle reports per segment in a 20-second period, the vehicle data needs to be aggregated before it is joined with the fixed-sensor data; further, the vehicle data is known to be noisy and must be cleaned before being processed. After the vehicle data is cleaned and aggregated, the fixed-sensor stream is (outer) joined with the aggregated vehicle readings, so that an aggregated vehicle reading joins with the corresponding fixed-sensor reading if the sensor reports speed < 45 MPH; in addition, all fixed-sensor readings are included in the join result. In this scenario, data cleaning and aggregation are done for all vehicle readings regardless of segment congestion status—a waste of resources. Communicating the uncongested segments to the aggregation and data-cleaning operators could avoid data cleaning and aggregation on those segments. Note this information is a two-dimensional specification with both spatial (segment) and temporal (time of congestion) extent. Furthermore, identifying the set of tuples that do not need to be processed depends directly on the incoming data; observations of the fixed-sensor data are required to determine which segments are congested. To allocate resources effectively, information about the importance of tuples must be communicated to the appropriate operators: the aggregate and data-cleaning operators in this example. In our solution, the join operator sends feedback punctuation, counter to the stream flow, to communicate with the aggregate and data cleaning operators so those operators may suppress the processing of vehicle readings from uncongested segments and as a result avoid unnecessary work.

In contrast to existing solutions for resource utilization, ours is a distributed solution. Such a solution is better for distributed data-stream systems, which we expect to be common in the future. Further, localizing both discovery of processing opportunities and exploitation of received feedback leverages the fact that operators are experts of their own domain and also avoids potentially expensive data transfers to a centralized optimizer.

The paper is organized as follows: In Section 2 we introduce the foundations of our inter-operator feedback model. In Section 3 we detail several sources and types of feedback one may find useful in a Data-Stream Management System (DSMS). Section 4 contains notions of correctness and several operator characterizations for supporting one type of feedback. In Section 5, we present an overview of the NiagaraST DSMS, highlighting architectural aspects relevant to our feedback strategies. In Section 6 we report experimental observations from an initial implementation of feedback support in NiagaraST. We discuss related work in Section 7. Section 8 contains final remarks and a summary of future work.

## 2. FEEDBACK OVERVIEW

Feedback punctuation is a mechanism for propagating information "backwards" through the runtime query plan, against the flow of data, for the purpose of dynamically adapting query processing to changing data, system, and client characteristics. Due to the dynamic nature of data streams, we anticipate that optimization opportunities will often be state-dependent and will be discovered during query processing. Feedback exploits those opportunities to increase system performance. We proceed to present four additional examples that illustrate potential uses of feedback and then discuss architectural considerations for the implementation of feedback.

**Example 1.** We add two extensions to the example described in the Introduction. First, instead of just congested and uncongested segments, we note that there may be a congestion classification scheme (low, medium, high), with vehicle readings from highly congested segments marked as high-priority. Such a scenario requires a more sophisticated prioritization mechanism—beyond simply dropping a subset of tuples—and is an example of the need for prioritized processing. We anticipate that a form of feedback punctuation, called *desired punctuation*, could support such prioritization. Second, in many situations, there may be additional, possibly user-specified, real-time information that needs to be considered when deciding which tuples to process; an example is a desire to receive as soon as possible data from highly-congested freeway segments. We anticipate feedback supporting such a need.

**Example 2.** Consider a slide-by-tuple window of range n, and windows $w_1, w_2, …, w_k$. Assume it is discovered that windows $w_3$ and $w_4$ are not required for the query result. In a scenario similar to Figure 1 (b), avoiding the processing of these windows by placing a filter at the bottom of the plan to filter out the tuples that belong to $w_3$ and $w_4$ is incorrect: Those tuples can be part of other windows. Thus selection filters are not sufficient for this example. All tuples may still need to be cleaned, but the aggregate can avoid working on the unnecessary windows.

**Example 3.** Consider the input stream of sensor data from the introduction, where sensors experience intermittent failures that cause them to report null values. Input tuples are filtered and split into two disjoint streams: clean and dirty. An operator called IMPUTE processes tuples for the dirty stream and uses an expensive method to replace the missing values with acceptable estimates. A specialized UNION operator (called PACE) performs the union of imputed and clean tuples, but bounds the maximum delay between clean and imputed tuples by ignoring tuples that are too late. When PACE detects that the maximum delay between clean and imputed tuples is being exceeded, it can generate feedback so that IMPUTE can avoid processing tuples whose timestamp is already too late with respect to PACE's current high-watermark of the timestamps seen, thus avoiding unnecessary work. An evaluation of this strategy is discussed in Section 6.

**Example 4.** In some data-stream systems, results are automatically produced periodically. In others, results may be produced in a poll-based fashion, with a user or application requesting results from the stream system when results are desired. Feedback can be used to support such on-demand result production by propagating the result request through the query tree so that results can be produced as needed and so that results do not have to be produced when not needed.

**Architectural Considerations:** One approach to supporting feedback consists of developing a centralized mechanism that actively monitors the performance of each operator, and tries to identify optimization or adaptation opportunities. This approach is illustrated in Figure 2 (a). Such a centralized monitor may also respond to application feedback. We are leery of a centralized solution because the centralized monitor might need access to the data stream itself as well as knowledge of the data-cleaning and aggregate-operator semantics. As shown in the examples, optimization or prioritization decisions are often state-dependent, thus the potential need for the centralized monitor to access the data stream. Transferring data to the centralized optimizer would be potentially very expensive in a distributed system. In addition, such a centralized optimizer would need to contain all possible strategies for discovery and exploitation of feedback for all operators and would require knowledge of all possible interactions across operators. In contrast, we propose a localized optimization approach, shown in Figure 2 (b). In this model, each operator locally implements discovery and exploitation mechanisms in terms of the operator's internal logic, input, and output schemas. Since feedback will be only propagated to the antecedent operators, the local design need not consider interaction with other operators. This model exploits the fact that an operator is an expert about its own domain. The localized approach extends well to distributed data-stream systems where communication between entities is expensive. The localized approach does require communication between operators, which may complicate system implementation. However, it may well be the case that the intermediary operators (i.e. the join in Figure 2) through which the punctuation is propagated know best about how to interpret and take advantage of the feedback information. As will be discussed in Section 5, NiagaraST already supports inter-operator message transmission and as such is amenable to the implementation of feedback punctuation.

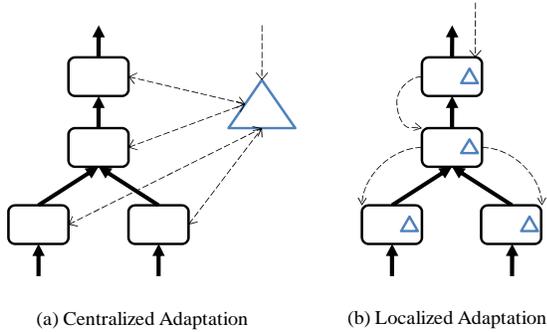

(a) Centralized Adaptation      (b) Localized Adaptation

**Figure 2. Centralized Adaptation vs. Localized Adaptation. Thick arrows indicate stream flow. Dashed arrows indicate feedback communication.**

## 3. SOURCES AND TYPES OF FEEDBACK

The motivating examples described in the Introduction and in Section 2 lead us to observe that (1) discovery of optimization opportunities generally occurs downstream in the query plan, and (2) propagating feedback information upstream may benefit query processing efficiency by having antecedent operators exploit the discovered opportunities. In this section, we present different types of feedback punctuations and their potential applications, discuss potential sources of feedback, and present a range of responses that operators may exhibit when receiving feedback punctuations. We begin with a brief description of punctuation.

### 3.1 Punctuation

Punctuation is a mechanism to unblock blocking operators and to purge state from stateful operators in data-stream systems [12]. Punctuation was originally implemented in the NiagaraST system and has since been implemented in Gigascope [7]. In general, punctuation is a method for communicating information about stream state between operators and as described in the literature [13] and as used in NiagaraST's Out-of-Order processing (OOP) architecture [9], provides information about the completion of substreams. For example, a punctuation may be used to indicate that no tuples with timestamps before '2008-12-08 9:00 AM' will be seen again in the data stream. An aggregate operator can use this information to produce results for and purge state from time periods (windows ending) prior to '2008-12-08 9:00 AM'.

Punctuations, as previously proposed, flow in the data stream and are represented similarly to tuples. Given a schema with three attributes, the third of which is the timestamp, the above-described punctuation can be denoted [*, *, ≤'2008-12-08 9:00 AM'], with the wildcard "*" indicating "any value". For the rest of this paper, we use the term *embedded punctuation* to refer to such punctuation that flows with the data stream.

### 3.2 Feedback Punctuation

Similar to *embedded punctuation* [12][13], feedback punctuation also carries predicates that *describe* the set of tuples associated with the feedback, but flows against the stream direction and carries an additional piece of information: the *intent* of the feedback. Feedback punctuation is not part of the stream. The "intent" that feedback punctuation carries does suggest a specific response that the receiving operator may want to perform.

### 3.3 Sources and Applications of Feedback

We have identified three kinds of feedback sources, as well as possible applications of feedback depending on the feedback's origin. The following list is not exhaustive.

**Explicit.** The definition of the query plan may include explicit policies that require enforcement, as in Example 3. Consider uniting two arbitrary streams, and a policy restriction mandated by an application that requires that the result stream exhibit no more than one minute of disorder relative to the tuple timestamps. The query and policy can be expressed in a SQL-like language as:

```
SELECT *
FROM stream1 UNION stream2
WITH PACE ON MAX(stream1.time, stream2.time)
     1 MINUTE
```

One possible implementation of such a restriction may translate the expression into a query plan with a PACE operator at the top of the query; PACE will compare the timestamp attributes of `stream1` and `stream2`. If tuples are lagging beyond the one minute tolerance with respect to the current high-watermark of the timestamps, PACE could generate feedback punctuation to inform antecedent operators that tuples with late timestamps have been seen and are being ignored, so production of such tuples should be avoided.

**Adaptive.** We envision that adaptive versions of existing operators may be able to discover processing opportunities in their streams. Recall the vehicle data stream and the fixed-sensor data stream from Section 1. Both streams are joined on location using tumbling windows of 1 minute. Assume two punctuations arrive consecutively on the vehicle stream: The first punctuation indicates that all vehicle data has been seen for window 3, and a second punctuation indicates that all vehicle data has been seen for window 4. If tuples are arriving in order, one can notice that no tuples in the sensor stream in window 4 will join with probe tuples, as window 4 is empty in the probe stream. A thrifty version of JOIN — THRIFTY JOIN — could detect that window 4 is empty, and provide feedback to the sensor stream. Antecedent operators in the sensor stream can choose to stop producing tuples that would be part of the useless window.

**Event-driven.** In addition to explicit declaration of policies and adaptive generation of feedback, we have considered event-driven feedback. Consider zooming into a section of the speed map. Feedback could be sent through the query to temporarily avoid processing tuples that pertain to areas of the speed map that are not visible.

### 3.4 Types of Feedback Punctuations

Embedded punctuations are used to trigger result production and state elimination in operators. Feedback punctuations have additional activities associated with their use, such as processing adaptations. In this section we discuss three types of feedback punctuations that we have identified. Feedback punctuations carry an *intent* (their type) and a *description* of the subset of interest (the predicate).

As typically used in practice, embedded punctuations punctuate a timestamp attribute and provide an indication of stream progress as described in the example in Section 3.1. In contrast, feedback punctuations typically punctuate a greater variety of attributes. Consider a schema (timestamp, datavalue), the data subset informally described as "all data before 10:00:00 AM" is represented as [≤'10:00:00 A.M.',*]. The subset "all tuples with data values greater than or equal to 50" is represented as [*, ≥50].

**Assumed.** *Assumed punctuation* has the intent of communicating a set of tuples to be avoided. Assumed punctuations are denoted by prepending a "¬" to a punctuation specification. Consider an operator, O, that produces feedback stating that any tuples with timestamp prior to 10:00:00 A.M. will be ignored – i.e., the issuer operator has determined that the described subset is of no further use and proceeds execution assuming that subset will no longer be seen. Consider an input schema (timestamp, datavalue). O's intention is to inform antecedent operators that its processing will continue as if it had received the embedded punctuation [≤'10:00:00 A.M.',*]. Assumed punctuations have more of the flavor of a hint than a command and is the type of feedback needed by Example 3 in Section 2.

**Desired.** *Desired punctuation* has the intent of prioritizing production of the set of tuples described. The issuing operator has determined there is value in receiving a subset of the data as soon as possible. Consider a scenario in which prioritized processing of subsets of tuples is required: A user drives through a freeway system and is interested in receiving the most up-to-date information available for her current location, but is willing to tolerate delay in data from other locations. Desired punctuation is a type of feedback that enables prioritization. In other cases, desired punctuation may be a preamble to assumed punctuation, serving as a "warning". Desired punctuations are denoted by prepending a "*?*" to a punctuation specification.

To illustrate the use of desired punctuation, consider a new binary operator: IMPATIENT JOIN. This operator is eager to produce results. Consider joining vehicle data and sensor data. Potentially, there are considerably fewer vehicles than sensors, as ad-hoc vehicles tend to be an expensive means to collect data. IMPATIENT JOIN can send desired punctuation feedback to the sensor stream saying "I have vehicle data for segment #3 and time period #7", expressed as ?[7,3,*] under schema (period, segment, data). Antecedent operators may prioritize processing data for that segment and time period, since IMPATIENT JOIN can use such results to produce output. Unlike assumed punctuation, this new type of feedback does not change the overall result of the issuing operator, but affects what the receiving operator may do, namely, the production time and order of its result stream.

**Demanded.** *Demanded punctuation* is the conceptual intersection of assumed and desired punctuation. The communicated message carries the sentiment of "I need this subset now". The issuing operator has detected it is more useful to receive a partial answer as soon as possible as opposed to a full answer too late. Like assumed punctuation, production of this type of feedback may be triggered by a utility-policy violation. Demanded punctuation carries the "warning" sense of desired punctuation, but unlike assumed and desired, demanded conveys the intent of being willing to accept an approximate result. Demanded punctuations are denoted by prepending a "*!*" to a punctuation specification.

For example, consider a financial speculator, whose margin of action is limited to a few seconds. The speculator needs to decide whether to buy or pass on a particular currency based on fluctuating exchange rates. She would like to receive a best guess estimate on the trend in the exchange rate in less than 5 seconds. A demanded punctuation may cause some aggregates to unblock and produce partial results. In this example, partial results are better than no results, or seeing results after the end of the margin of action.

### 3.5 Range of Operator Responses to Feedback

Feedback punctuations enable a range of responses by the receiving operators, including exploiting a processing opportunity or propagating the received feedback to its antecedent operator(s). In this section, we discuss a subset of these responses relevant to assumed punctuations.

Consider the stateful operator AVERAGE whose input stream is a set of speed readings from probe vehicles and which aggregates those speed readings into one-minute averages; that is, AVERAGE groups all speed readings in a one-minute period (say 9:00-9:01) and outputs an average speed for that minute. AVERAGE has input schema (timestamp, speed) and output of the form (minute, avg(speed)). Consider AVERAGE receiving an assumed punctuation of the form ¬[<'2008-12-08 9:01',*]. Based on this punctuation, AVERAGE can avoid production of results for minutes earlier than '2008-12-08 09:01', perhaps simply by avoiding producing the result once it is computed, or by more aggressively purging its internal state and avoiding recreation of windows known to be unneeded.

Now consider AVERAGE receiving an assumed punctuation of the form ¬[*,≥50], indicating that windowed averages of speeds greater than or equal to 50 will be ignored. Suppressing active windows is not a correct response for this feedback. Suppose window 4 is active and has a current partial average of 51. Purging this window from the hash table would be a mistake, as future tuples arriving on its input could cause the average for window 4 to drop below 50. Similarly, propagating assumed punctuation is an incorrect response, as no assumptions on unseen tuples can be made. Correct response options that optimize processing exist. If embedded punctuation arrives on its input and indicates that all tuples for window 4 have been seen, and the current partial average of that window is 51, AVERAGE can avoid constructing and producing an output tuple.

The range of appropriate responses can depend on operator state and nature. Consider a MAX operator on the speed values from the probe stream, whose output is of the form (minute, max(speed)). The MAX operator maintains a partial aggregate per active window. If MAX receives an assumed punctuation of the form ¬[*,≥50], MAX has the opportunity to perform at least two actions: It should close all open windows for which the partial aggregate matches the assumed punctuation, as the current aggregate already satisfies the predicate. It should also prevent windows that match the assumed punctuation from forming. Prevention can be locally enforced with a *guard* on MAX's input, since antecedent operators may still produce tuples leading to undesired windows before they receive and act on propagated

feedback – for example, receiving a tuple with value 40 would create a new incorrect partial aggregate. Other aggregate operators may have different response options because of their nature (i.e., COUNT's produced result increases monotonically, SUM's doesn't.)

The range of correct responses to a feedback punctuation may also be limited by the contents of the punctuation. Consider probe-vehicle and traffic-sensor streams, with schemas `detector(id, freeway_id, milepost, timestamp, speed)` and `probe(id, freeway_id, milepost, timestamp, speed)`. Consider joining the streams on `freeway_id`, `milepost`, and `timestamp`, leading to an output schema (`probe.id, freeway_id, milepost, timestamp, probe.speed, detector.id, detector.speed`). If JOIN receives the assumed punctuation ¬[*,11,30,≤'10:00:00 A.M.',*,*,*] it can safely propagate the feedback to both of its inputs. In contrast, if JOIN receives the assumed punctuation ¬[*,*,*,*,≥50,*,*], feedback can only be propagated to the antecedent operators of the probe stream, as there is no attribute in the detector speed that matches the feedback schema.

## 4. NOTIONS OF CORRECTNESS

While we have identified what operators may discover and propagate as feedback; however, a critical component of the feedback approach consists in defining what it means for operators to exploit processing opportunities expressed by received feedback. The ranges of responses can include guarding input and output, purging state for assumed punctuation, internal priorization of subsets of tuples for desired punctuation and immediate production of partial results for demanded punctuation. Our initial exploration has focused on assumed punctuation for which we have defined notions of correct exploitation and safe propagation. In this section we present these notions, as well as a framework to characterize operator responses. We also discuss some concerns about feedback exploitation semantics, in particular, expiration of feedback information.

The following terminology is used throughout the remainder of this section. O is an operator which consumes an input stream $S_I$ and produces an output stream $S_R$. O receives assumed feedback punctuation f, and produces assumed feedback punctuation g. The expression `subset(stream, punctuation)` refers to the set of tuples in stream that match the predicate `punctuation`.

### 4.1 Correct Exploitation
We begin by defining correct exploitation of an assumed feedback punctuation.

**Definition 1.** An operator O *correctly exploits* a processing opportunity expressed by assumed punctuation f if and only if, upon exploitation, O produces an output stream S such that $S_R$ – `subset(stream, punctuation)` ⊆ S ⊆ $S_R$.

By bounding correctness, we allow operators to exhibit a null response (S ≡ $S_R$) and still deem that response as correct. In contrast, other operators may aim for maximum exploitation, that is producing $S_R$ – `subset(stream, punctuation)`. No exploitation of assumed punctuation should insert tuples in S that would not have normally appeared in $S_R$. Some operators, such as DUPLICATE, have more than one output stream. In the case of DUPLICATE, the operator's definition implies both output streams need to be identical, hence exploiting an opportunity would either affect either both outputs or none.

This definition of correctness allows for reasoning only about a particular operator, not the entire query result. In practice, aspects that require consideration for overall analysis include in-flight tuples, feedback propagation delay, and operator scheduling.

### 4.2 Safe propagation
Feedback may be propagated to antecedent operators. Such propagation depends on the ability to compute a function that maps from output to input schema, which may not exist for all attributes. Even when the mapping exists, special care must be taken in the construction of the punctuation to be propagated to antecedent operators.

**Definition 2.** An operator O *safely propagates* g if any antecedent operator's exploitation of the opportunity expressed by g does not alter O's correct exploitation of g.

Consider two streams with the following schemas: `A(a, t, id)` and `B(t, id, b)`. Consider an equi-join of both streams on t and id. The output schema of JOIN is `C(a, t, id, b)`. For feedback f ¬[*,3,4,*], JOIN can safely propagate ¬[*,3,4] and ¬[3,4,*] to inputs A and B, respectively. If f is ¬[50,*,*,*], the only possible correct propagation is ¬[50,*,*] to input stream A. For the feedback f ¬[50,*,*,50] no safe propagation exists; propagating ¬[50,*,*] and ¬[*,*,50] to both inputs may incorrectly suppress production of a tuple not covered by the feedback punctuation, such as the tuple <49,2,3,50>.

### 4.3 Operator Characterization
A central component of our architecture consists in extending the design of operators to respond to assumed punctuation. Since the intention of said feedback is to avoid the processing of the subset of tuples expressed by the predicate, operators need to have a menu of actions to perform.

Some general strategies we have identified include (1) activating a guard on the output (avoid emitting a tuple that matches the feedback), (2) activating a guard on the input (avoid computation on a tuple that matches the feedback), and (3) purging internal state of tuples that match the feedback. Each operator's design has to carefully consider these options and the combination of both to guarantee correct exploitation.

To characterize operators, it is useful to *partition* their output schemas in several sets that help distinguish attributes of interest, computed attributes, or carried attributes. Some operators can be characterized without partitioning.

To illustrate operator characterization, consider the window aggregate operator COUNT. We can partition its output schema as (g, a) where g is the set of attributes used for grouping, and a is the count. Table 1 formalizes several cases of assumed feedback punctuation that may be received by COUNT.

Operator characterization requires reasoning about the nature of the operator and guaranteeing that both exploit and propagation rules, if enacted, are correct and safe respectively. An example of an incorrect response to feedback of the form ¬[g,*] would be to simply purge local state without guarding the input, as incoming tuples may recreate the undesired group. Some responses are limited to one possible action. For example, the response to [*,a] is limited to only guarding the output, since count can monotonically increase and a purge upon reception of feedback produces an incorrect result.

**Table 1. A Characterization for COUNT.**

| Punctuation | Local exploit | Propagation |
|---|---|---|
| ¬[g,*] | 1. Remove group g from local state<br>2. Guard input (g) | 1. Propagate g (in terms of input schema) |
| ¬[*,a] | 1. Guard output (a) | |
| ¬[*,≥a]<br>¬[*,>a] | 1. G ← ids in local state that match the predicate<br>2. Purge state (G)<br>3. Guard input (G) | 1. Propagate G (in terms of input schema) |
| ¬[*,≤a]<br>¬[*,<a] | 1. Guard output (≤a or <a) | |

Characterization of other operators follows a similar strategy: First, we find a meaningful partition of the output schema, and then cover exploitation and propagation.

A meaningful partition for JOIN considers three sets: the set of attributes unique to the left input stream L, the set of join attributes J, and the set of attributes unique to the right input stream R. JOIN's output schema is (L,J,R). We will refer to the schema of the inputs as (L,J) and (J,R). The following characterization is not exhaustive—it only expresses exact values received for attributes. We also use loosely lower case variables to denote received variable—which may be values over a subset of the sets of interest.

Some operators require less analysis: SELECT, for example, maintains no internal state, and assumed punctuation can simply be added to its select condition. Ad-hoc operators, such as IMPUTE, are easily specialized. The characterization of stateful, general-purpose operators requires careful analysis.

### 4.4 Feedback Implementation Issues

State accumulation is undesirable in DSMSs. Keeping track of enacted feedback actions may entail state accumulation (not of tuple data, but of predicates). This suggests that feedback information may be best implemented and supported when there are *guarantees* that no unnecessary accumulation will occur.

Feedback punctuations must be subordinate to what the stream can represent and bound—a notion analogous to punctuation schemes [14]. Punctuation schemes help determine which queries can be executed over punctuated streams, by providing a framework to identify if a particular punctuation scheme benefits the whole query by freeing state and unblocking operators.

We have found examples of both supportable and unsupportable feedback. Supportable feedback examples have a common characteristic: they occur on *delimited* attributes. We consider an attribute to be delimited if it is covered by embedded punctuation. For example, if the stream contains punctuation on a timestamp attribute, the timestamp attribute is considered delimited. Consider a bid-auction stream. Feedback of the type "Do not show bids prior to 1:00 p.m." is easily supported. The progressive nature of the timestamps guarantees that eventually we will see a punctuation indicating all data elements up to 1:00 pm have been seen—at which point operators can free feedback-related guards and state. Similarly, state associated with feedback expressing "Do not produce results related to bidder #2 for auction #4" will be cleansed when auction #4 finishes, or when punctuation arrives indicating that we have seen all bids from bidder #2. In contrast, feedback of the type "Don't show bids more than $1.00" will leave state in the operators, as it is unlikely to punctuate on amounts. In this case, the user should have issued a different query, as the overall result has changed.

It is important to distinguish from available correct actions and supported actions. The exploit responses for JOIN (Table 2) could be dangerous if internal state is cleansed and then there is a request to ignore the previously sent feedback. Our current model assumes there are no retractions on feedback, that is, once enacted it is final and adheres to correctness. This requirement can be relaxed, but will limit the range of correct exploitation actions. JOIN, for example, would be limited to guard its output.

## 5. NIAGARAST ARCHITECTURE

NiagaraST is a stream processing system that has as its foundation the publicly-available Niagara Internet Query System developed at the University of Wisconsin-Madison [10]. NiagaraST inherits the general structure of its architecture from Niagara; however, many modifications were made to convert Niagara from its origins as an Internet query system to the NiagaraST stream processing system. These changes included modifications to support stream query processing as well as performance improvements. We focus on the query execution architecture of NiagaraST as that portion of the system is the most unique and most applicable to data stream processing.

**Table 2. A Characterization for JOIN.**

| Punctuation | Local exploit | Propagation |
|---|---|---|
| ¬[*, j⊆J,*] | 3. Purge matching tuples from hash tables<br>4. Guard input | 2. Propagate ¬[*, j⊆J] to left input, and ¬[ j⊆J,*] to right input |
| ¬[l⊆L,*,*] | 2. Purge matching tuples from left hash table<br>3. Guard input | 1. Propagate ¬l⊆L,*] to left input |
| ¬[*,*, r⊆R] | 4. Purge matching tuples from right hash table<br>5. Guard input | 2. . Propagate ¬[*, l⊆L] to right input |
| ¬[l⊆L,*,r⊆R] | 2. Guard output | |

At a very high level, the NiagaraST architecture is what is often called a "push-based" or "pipelined" architecture. Operators run as threads connected by inter-operator queues of tuples. That is, operators send and receive data by writing to and reading from queues. Each operator is run in its own thread and thread scheduling is left to the operating system. This architecture is very amenable to stream processing since data flows through the operators in a pipelined fashion as is desired in a data stream system. Further, the architecture naturally supports inter-operator parallelism, which is essential for a stream system. In addition to the data queues connecting operators, NiagaraST also supports control messages that flow both up and down the query tree.

NiagaraST keeps the basic operator, queue and control message structure from Niagara but adds punctuation, windowed operators and OOP to support data stream processing and performance improvements to the inter-operator queues and the XML parser.

**OOP and Punctuation:** NiagaraST uses the Out-of-Order Processing (OOP) architecture for data-stream processing. This architecture separates stream progress from physical stream arrival properties. This separation allows order-agnostic operator implementations and provides for flexible and efficient stream query evaluation [9]. OOP is based on the WID technique for handling disorder [8] which uses punctuation to describe and communicate stream progress [12][13]. OOP is a scalable architecture, and is expected to work particularly well in distributed and parallel environments where its ability to easily accommodate disorder and will be required. Further, OOP's ability to support ordered as well as order-insensitive operator implementations provides a basis for stream query optimization.

**Inter-Operator Communication:** In NiagaraST the inter-operator queues consist of pages of tuples[1]. This structure enforces batching of tuples, so that each operator must produce a page of tuples before the subsequent operator is allowed to process these tuples. Such batching limits the context switching between operators improving system performance. Other researchers have also observed the value of batching tuples [2]. A disadvantage of requiring tuples to be produced in batches, is that a slow stream may take a long time to produce a page of tuples. In NiagaraST, this issue is resolved by having punctuations flush pages; that is, a page is flushed to the queue (and sent to the subsequent operator) when either the page is full or when a punctuation is written to the page.

A relatively unique feature of Niagara and NiagaraST, is support for control messages that flow both directions in the operator tree. Control messages are out-of-band communication and are given high-priority and processed before pending tuples. In NiagaraST, control messages carry messages such as end of include stream and shutdown in the downstream (with data flow) direction and include feedback punctuation and shutdown messages in the upstream (against data flow) direction.

Figure 3 shows a schematic of inter-operator connections in NiagaraST. As shown in this figure, pages of tuples (denoted as blue boxes) and downstream punctuation (denoted as red boxes) flow from Operator1 to Operator2.

**Operator Control:** Operator control in NiagaraST is complex due to the fact that the operators must send and receive data and out-of-band control messages. The control is made simpler by the fact that each operator is a thread. As such, each operator has an object that it sleeps on when it has no work to do. An operator is awakened when a new data page or control message is sent to it.

---
[1] The NiagaraST team thanks Josef Burger for his suggestions regarding inter-operator queue design.

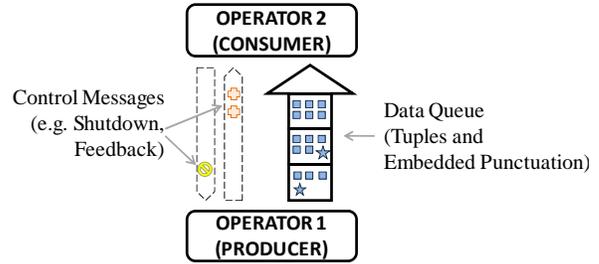

**Figure 3. NiagaraST Inter-Operator Connections**

**XML Parser:** The input data format of Niagara and subsequently NiagaraST is XML. The original XML parser was a traditional DOM parser which parsed XML documents into tuples that consisted of and tuples consisted of arrays of pointers to DOM elements. The parsing was too slow, therefore in NiagaraST, we implemented a parser based on a XML SAX parsing which uses SAX events to parse XML documents. The input is stored as pages of SAX events; DOM structures are materialized only as they are needed. The addition of this new "SAXDOM" parser resulted in major performance improvements.

**Feedback Support:** Our initial implementation of feedback in NiagaraST exploits the use of the control channel. In NiagaraST, control messages have two fields: a message type, such as shutdown or end-of-stream, and the control message. We added a new type to support feedback (assumed), and serialize the punctuation as a control message. Receiving operators are made feedback aware by adding processing actions. Upon receiving a control message, if said message is assumed punctuation, the control message is interpreted and acted upon. If mapping and further propagation are possible, a new control message is created and put in the upstream control channel. Feedback unaware operators ignore feedback and are unable to further propagate it.

## 6. EXPERIMENTAL OBSERVATIONS

We implemented support for assumed feedback punctuation in NiagaraST and tested two scenarios amicable to assumed punctuation. We conducted the experiments on a 2.8 GHz Pentium 4 machine with 1GB of RAM, running Windows XP.

**Experiment 1.** An implementation of the scenario described in Example 3, in which a stream processor filters an incoming stream of traffic sensor data and determines which tuples require imputation due to missing data. In this case, IMPUTE performs an archival lookup of similar tuples to produce an estimate. For each tuple that requires imputation, one database query is issued. We induced an extreme case in which tuples that require imputation alternate with non-imputed tuples in the stream. The query plan is shown in Figure 4 (a). This query plan splits incoming data into two disjoint streams, one containing tuples that require imputation and the other with tuples that can be output immediately.

When no feedback is used in the Imputation Query Plan (i.e., PACE is simply UNION), there is a considerable divergence in arrival time between imputed and non-imputed tuples, due to the fact that imputed tuples are much more expensive to produce than clean tuples. The divergence is illustrated in Figure 5. As time passes, the imputed tuples become more and more delayed with respect to the non-imputed tuples. Once the imputed tuples are delayed beyond a threshold, they are useless as the speed map must be produced in real time. For this scenario, a measure of improvement consists in the number of timely tuples that appear in the query result—that is, no later than a tolerance parameter specified in PACE. When PACE detects that the divergence in timestamps between clean and imputed tuples exceeds a tolerance parameter, it produces assumed punctuation informing that tuples with timestamps less than the current high watermark are no longer needed. This feedback causes IMPUTE to purge tuples that are already late from its current state, and to process more recent tuples. The effect of this use of assumed punctuation is shown in Figure 6. Our experimental runs used 5000 tuples and showed that only 29% of imputed tuples are dropped when PACE and feedback punctuation are used (Figure 6), as opposed to 97% of imputed tuples that arrived beyond the tolerated divergence when PACE and feedback punctuation are not used (Figure 5).

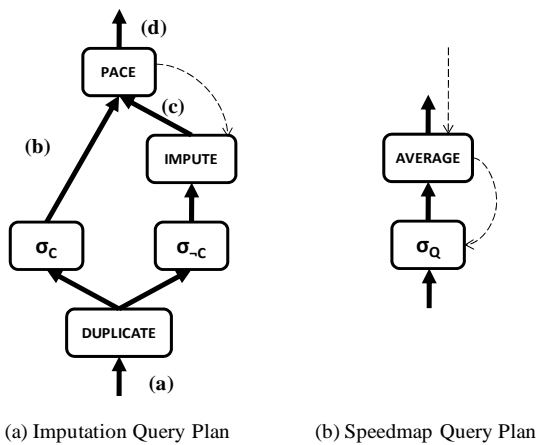

(a) Imputation Query Plan  (b) Speedmap Query Plan

**Figure 4. Query plans for Experiment 1 and Experiment 2.**

**Experiment 2.** A speed map that represents the status of a freeway system is displayed in a user's on-board navigation system. The user zooms into different segments of the map. The query plan that serves this application has a data quality filter at the bottom of the query. Assumed punctuation is sent through the query plan to avoid processing tuples that match the segments not displayed by the user (hence saving processing cost.) The query plan is shown in Figure 4 (b). This experiment is intentionally simple to illustrate savings in processing undesired tuples – we are not suggesting general purpose speed maps operate this way.

For Experiment 2, we used three different optimization schemes (F1 – F3 in Figure 7). Scheme F1 mounts a guard on the output of AVERAGE. Scheme F2 is more aggressive, as it avoids averaging groups known to be of no interest to the user. Scheme F3 further propagates the feedback to the quality filter. In this experiment, we simulated a traffic stream of 18 hours of data at a 20 second resolution, with 9 freeway segments and 40 detectors per segment (~1 million tuples). We used total query execution time as the metric. We also tested the frequency at which feedback is sent, assuming the vehicle viewing the map switches segments every 2, 4, or 6 minutes. We observed no discernible overhead as the frequency of feedback increases. Results are shown in Figure 7, where F0 is the baseline execution without any feedback. Once feedback is propagated to AVERAGE (Scheme F1), query execution time drops 50%. Scheme F2 further reduces processing time by 61%, while Scheme F3 achieves a 65% gain.

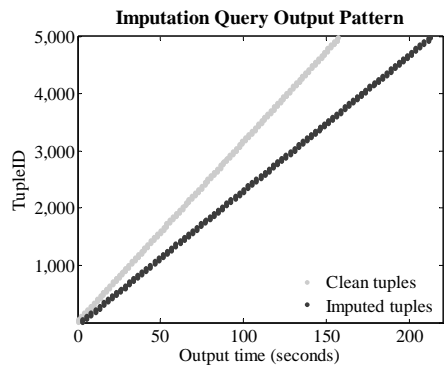

**Figure 5. Imputation query plan without feedback.**

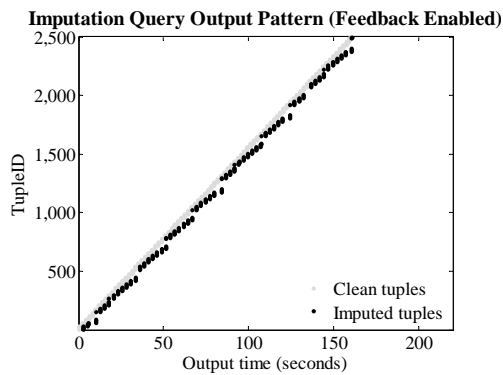

**Figure 6. Imputation query plan with feedback.**

## 7. RELATED WORK

Feedback punctuations flow against the stream direction, and carry an intent, as opposed to embedded punctuations [12][13]. Performance optimizations based on statistics and feedback have been discussed in the context of DSMSs. TelegraphCQ allows users to specify priorities during execution [5]. TelegraphCQ relies on a query processing mechanism called *Eddy*, which continuously reorganizes operators in a query plan depending on learned statistics, such as selectivity of operators [1]. Aurora enforces QoS via a centralized Load Shedder. Borealis, the distributed version of Aurora, introduces the concept of Control Lines [4]. These lines feed operators with revised parameters that affect the operator's behavior. Borealis maintains a centralized repository of applicable functions. An operator called Bind analyzes the incoming stream to select (or create) a new function from the repository. The results of Bind are sent through the control line to the operator, along with instructions on how to apply the changes. An application of feedback discussed in Borealis consists in enabling online modification of continuous queries [2]. StreaMon is a component of the STREAM system that profiles and adapts query plan parameters based on collected statistics—a performance-improving feedback loop [3]. CEDR is a data stream and event processing system under development at Microsoft Research. CEDR's operators produce speculative (i.e. partial) results followed by subsequent correctness guarantees [6]. AT&T's Gigascope DSMS includes a type of punctuations—called heartbeats—that signal the passing of time, but their work does not mention feedback mechanisms [7]. Punctuation uses continue to expand beyond delimiting epochs in the stream; Nehme et al. have used notions of punctuations as security constraints to enact access control policies [11].

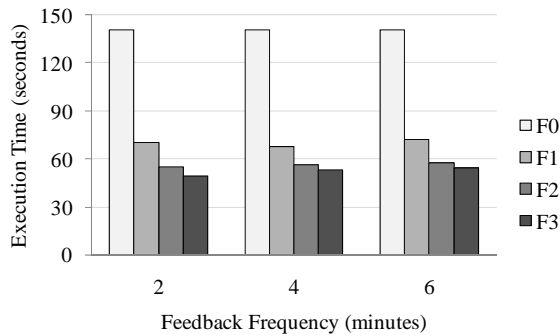

**Figure 7. Comparison of Feedback Schemes**

DSMSs have focused on collecting and distributing feedback information by statically placing monitors in the query plan and directly sending parameter changes to operators. Our approach differs significantly in at least two aspects: (1) the use of punctuation to convey the feedback messages, and (2) giving operators the ability to create, consume, and propagate feedback, eliminating the need for centralized managers.

## 8. CONCLUSIONS AND FUTURE WORK

We have presented a feedback scheme to enable query processing strategies using punctuations that describe subsets of interest of the data stream. Our mechanism consists of three types of feedback: assumed, demanded, and desired, and recognizes that processing opportunities may be discovered dynamically by operators, responding to either application-time specifications or changes in data (such as data bursts). By localizing discovery, processing, and exploitation, our approach should scale well to distributed systems, as communication between operators only occurs between subsequent and antecedent operators. The Inter-Operator Connection architecture in NiagaraST suggests a template for implementation of this mechanism in other systems. We have provided initial notions of correctness for assumed feedback punctuation, and examples of operator design choices.

Future work will explore implementation of feedback support on other operators, and add theoretical descriptions of correct exploitation and safe propagation for desired and demanded punctuation. A more general aspect of interest is the time to live of feedback. Our current notion of allowable feedback recognizes that embedded punctuation may delimit the range of supported feedback in a given query.

## 9. ACKNOWLEDGMENTS

The authors wish to thank David Maier, Robert Bertini, Vassilis Papadimos, and Emerson Murphy-Hill for advice, support, and discussion. This work is supported by NSF Awards (IIS 00-86002, IIS 06-12311) and NSF CAREER Award (0236567), OTREC (grant 2007-64), and CONACyT México (178258).